# Quadrupolar Kondo Effect in Non-Kramers Doublet System PrInAg$_2$

**Osamu SUZUKI, Hiroyuki S. SUZUKI, Hideaki KITAZAWA and Giyuu KIDO**

*National Institute for Materials Science,*

*1-2-1 Sengen, Tsukuba 305-0047, Japan*

**Takafumi UENO, Takashi YAMAGUCHI, Yuichi NEMOTO and Terutaka GOTO**

*Graduate School of Science and Technology, Niigata University*

*2-8050 Igarashi, Niigata 950-2181, Japan*

**Keywords:** PrInAg$_2$, quadrupolar Kondo effect, elastic constant, non-Kramers doublet

We performed ultrasonic measurement on the rare-earth intermetallic compound PrInAg$_2$ to examine the quadrupolar Kondo effect associated with the non-Kramers $\Gamma_3$ doublet ground state. The characteristic softening of the elastic constant $(c_{11}-c_{12})/2$ below 10 K in PrInAg$_2$ is attributed to a Curie term in quadrupolar susceptibility for the quadrupole $O_2^2 = J_x^2 - J_y^2$ of the stable $\Gamma_3$ ground state. $(c_{11}-c_{12})/2$ turns to a slight increase with the $-\ln T$ dependence below 0.1 K, which suggests the quenching of the quadrupolar moment in the quadrupolar Kondo state. Under applied magnetic fields of 10 T and 15 T above 8.7 T corresponding to the Kondo temperature $T_K$ of ~ 0.86 K, the behavior of $(c_{11}-c_{12})/2$ is described in terms of quadrupolar susceptibility for the stable 4f$^2$ state.

In metallic compounds with magnetic ions, the antiferromagnetic interaction between the magnetic moment of a localized electron and the spin of conduction electrons leads to the Kondo effect induced by the quenching of the localized magnetic moment by the screening of conduction electrons.[1] In his milestone work, Kondo reported that the logarithmic divergence of resistivity is driven by the scattering of conduction electrons caused by magnetic impurities embedded in metal.[2] The Kondo lattice in Ce-based compounds induces the heavy-Fermion state, which is characterized by an enhanced Sommerfeld coefficient, a large Pauli paramagnetic susceptibility, and a low-temperature resistivity proportional to $T^2$. The Fermi liquid behavior is adopted for the Kondo state of spin 1/2 as well as the Kondo lattice in periodically arrayed Ce ions.[3]

The Kondo effect depends on the electronic properties of impurity ions and their interactions with the host metal. Nozières and Blandin discussed theoretically that the ground state depends on the number of scattering channels of conduction electrons ($N_{ch}$) and the magnitude of the localized impurity spin ($S$).[4] The compensation of impurity moments for $N_{ch}=2S$ leads to the Fermi liquid behavior. The overscreening for $N_{ch}>2S$ may lead to the non-Fermi-liquid (NFL) behavior. Cox proposed the quadrupolar Kondo effect to discuss the overscreening effect of the non-Kramers doublet ground state for $f^2$ state in U-based compounds of (Y, U)Pd$_3$.[5] Many theoretical investigations confirmed that the NFL state is relevant for the isolated non-Kramers doublet state of single impurities embedded in metal.[5-8] However, the quadrupolar Kondo effect in a periodically arrayed non-Kramers doublet in metal is still unresolved. Jarrell *et al.* examined the two-channel Kondo lattice leading the incoherent metal with the NFL properties.[9] Tsuruta *et al.* reported that the multichannel degenerate Anderson lattice may lead to the heavy-Fermion state as well as the NFL state.[10] The experimental clarification of the quadrupolar Kondo effect in Pr-based compounds with the non-Kramers $\Gamma_3$ doublet ground state is one of the most interesting issues in the strongly correlated electron physics.

A few systems of the Pr-based intermetallic compounds PrPb$_3$, PrPtBi and PrInAg$_2$ possess the non-Kramers $\Gamma_3$ doublet ground state.[11-13] PrPb$_3$ and PrPtBi exhibit quadrupolar ordering at low temperatures. Yatskar *et al.* reported that PrInAg$_2$ with a cubic Heusler-type crystal structure is an ideal system for examining the quadrupolar Kondo effect, because no phase transition has been found in this system down to 0.1 K.[14] Inelastic neutron scattering experiments on PrInAg$_2$ revealed crystal-field transitions indicating the non-Kramers doublet $\Gamma_3$ (0 K) for the ground state, and the excited states of $\Gamma_4$ (70 K), $\Gamma_5$ (96 K), and $\Gamma_1$ (176 K).[15] The temperature dependence of specific heat shows a clear Schottky anomaly at around 30 K. The electrical resistivity of PrInAg$_2$ shows dependence on $\Delta\rho \propto \log T$ below 1 K down to 50 mK, which is inconsistent with $\Delta\rho \propto \sqrt{T}$ anticipated for the single impurity quadrupolar Kondo model.[16] Furthermore, the specific heat shows a broad maximum at around 0.4 K and a huge specific heat coefficient $C/T = \gamma = 7$ J/mol K$^2$ below 0.1 K.[14, 17] These results are considered to be associated with the quadrupolar Kondo effect due to the strong correlation between conduction electrons and the periodically arrayed non-Kramers $\Gamma_3$ doublet in PrInAg$_2$. Recently, Kawae et al. have reported the low-temperature properties of the diluted compounds Pr$_x$La$_{1-x}$InAg$_2$ for $0 \leq x \leq 0.6$, where the logarithmic $T$ dependences of magnetic susceptibility and specific heat below $T = 15$ K are presented in connection with the NFL behavior of the quadrupolar Kondo effect.[18] Furthermore, the scaling of the specific heat coefficient $C/T$ below 10 K demonstrates a characteristic temperature of $T^* = 1$ K ~ 0.7 K for $x = 0.05$ ~ 0.6 for the on-site hybridization energy between the 4f electrons and conduction electrons.

Rare-earth compounds exhibit frequently elastic softening with decreasing temperature, which is described in terms of quadrupolar susceptibility for the localized 4f-electron system. The non-Kramers $\Gamma_3$ doublet with an SU(2) symmetry possesses three quantum degrees of freedom, two electric quadrupoles (i.e., $O_2^0 = \frac{1}{\sqrt{3}}\left(2J_z^2 - J_x^2 - J_y^2\right)$ and $O_2^2 = J_x^2 - J_y^2$) and one magnetic octupole (i.e., $T_{xyz} = \overline{J_x J_y J_z}$)[19]. Here, the bar means the sum of cyclic permutations on $x$, $y$, and $z$. The measurement of the elastic constant $(c_{11}-c_{12})/2$ is a suitable probe for observing the quadrupolar susceptibility for $O_2^2$ of the non-Kramers $\Gamma_3$ doublet.[11, 12, 20] The direct observation of the octupole $T_{xyz}$ is not realized, because appropriate experimental probe conjugated to the octupole has not been found thus far. The magnetic susceptibility measurement is insensitive for the present system because of the absence of a magnetic dipole for the non-Kramers $\Gamma_3$ doublet. In this letter, we present the elastic constant of a single-crystalline sample of PrInAg$_2$ as a function of temperature and magnetic field. The quenching of the quadrupolar moment at low temperatures is argued in connection with the quadrupolar Kondo effect in crystalline PrInAg$_2$.

The single-crystalline sample PrInAg$_2$ was synthesized by the Bridgman method in a closed Mo crucible. We cut the sample into rectangular pieces, typically 2.0 x 1.5 x 1.0 mm$^3$. For ultrasound velocity measurement, we used a homemade ultrasound apparatus based on the phase comparison technique, which can detect a relative sound velocity change $\Delta v/v_0 \sim 10^{-6}$. We used a transverse sound wave with a fundamental frequency of 15 MHz in ultrasound velocity measurement. We used thin piezoelectric LiNbO$_3$ plates with evaporated gold electrodes on both their surfaces as ultrasound generator and detector. A pair of transducers is glued on the parallel surface of the specimen by silicone rubber (Shin-etsu Silicone, Co., Ltd.). Elastic constant $(c_{11}-c_{12})/2$ is measured using a transverse sound wave propagating along the [110] axis and polarizing along the [1$\bar{1}$0] axis. The elastic constant $c$ can be estimated using formula $c = \rho v^2$, where $\rho = 8.843$ g/cm$^3$ is the mass density of PrInAg$_2$ and $v$ is the sound velocity. Low-temperature ultrasonic measurement was performed using a homemade $^3$He evaporation refrigerator down to 500 mK and a top-loading-type $^3$He-$^4$He dilution refrigerator down to 10 mK. These refrigerators were equipped with superconducting magnets.

As shown in Fig. 1(a), $(c_{11}-c_{12})/2$ in zero magnetic field increases with decreasing temperature from 120 K and exhibits a broad shoulder at about 40 K. This shoulder is considered to be due to the 4f electrons of the Pr ion because no anomaly was observed in $(c_{11}-c_{12})/2$ of the nonmagnetic reference compound LaInAg$_2$, which is not presented here. The plateau in Fig. 1(a) may have a dynamical effect due to the large oscillation amplitude of the Pr ion, because the specific heat of PrInAg$_2$ shows no anomaly at around this temperature. Below about 40 K, $(c_{11}-c_{12})/2$ increases slightly with lowering temperature and shows a maximum at about 10 K. With further lowering temperature, $(c_{11}-c_{12})/2$ shows a softening of about 3% below 10 K down to 0.1 K. On the other hand, $c_{44}$ shows a slight softening of about 0.1% below 5 K.

The softening of $(c_{11}-c_{12})/2$ below about 10 K is accounted for by the quadrupole-strain interaction $H_{QS} = -g_{\Gamma 3}(O_2^0\varepsilon_u + O_2^2\varepsilon_v)$. Here, the quadrupoles $O_2^0$ and $O_2^2$ couple to the elastic strains $\varepsilon_u = \frac{1}{\sqrt{3}}\left(2\varepsilon_z^2 - \varepsilon_x^2 - \varepsilon_y^2\right)$ and $\varepsilon_v = \varepsilon_{xx} - \varepsilon_{yy}$, respectively. $\varepsilon_v$ is relevant for the transverse $(c_{11}-c_{12})/2$ mode used in the ultrasonic measurement. $g_{\Gamma 3}$ is a coupling constant. Furthermore, we introduce the quadrupole intersite interaction as $H_{QQ} = -\Sigma_i\, g'_{\Gamma 3}\, (<O_2^0>O_2^0(i) + <O_2^2>O_2^2(i))$. Here, $g'_{\Gamma 3}$ denotes a coupling constant in mean field approximation. The elastic softening of $c = (c_{11}-c_{12})/2$ in Fig. 1 is reproduced well by

the formula[21]

$$c(T) = c^0 - \frac{N g_{\Gamma_3}^2 \chi_{\Gamma_3}(T)}{1 - g'_{\Gamma_3} \chi_{\Gamma_3}(T)}, \tag{1}$$

where, $N$ is the number of Pr ions in unit volume. $\chi_{\Gamma_3}(T)$ is the quadrupolar susceptibility for $O_2^2$ per Pr ion as

$$-g_{\Gamma_3}^2 \chi_{\Gamma_3} = \left\langle \frac{\partial^2 E_n}{\partial \varepsilon_v^2} \right\rangle - \frac{1}{k_B T}\left[ \left\langle \left( \frac{\partial E_n}{\partial \varepsilon_v} \right)^2 \right\rangle - \left\langle \frac{\partial E_n}{\partial \varepsilon_v} \right\rangle^2 \right], \tag{2}$$

where, $E_n$ is the $n$-th crystal-field energy perturbed by the Hamiltonian $H_{QS}$ up to the second order of $\varepsilon_v$. The one-site quadrupolar interaction $|g_{\Gamma_5}| = 5.13$ K and intersite quadrupolar interaction $g'_{\Gamma_5} = -0.004$ K are used for obtaining the calculation solid line in Fig. 1. The background part $c^0 = aT + b$ is denoted by a broken line, where $a = -0.0015 \times 10^{10}$ erg/cm$^3$ K and $b = 10.1701 \times 10^{10}$ erg/cm$^3$. The softening of $(c_{11}-c_{12})/2$ above 0.3 K is described well by the Curie term in the quadrupolar susceptibility $\chi_{\Gamma_3}(T)$ for $O_2^2$ associated with the stable 4f$^2$ shell with the non-Kramers $\Gamma_3$ doublet ground state.

As shown in the inset of Fig. 1 (a), however, the experimental result of $(c_{11}-c_{12})/2$ below about 0.3 K deviates from the calculation results based on the quadrupolar susceptibility of the Pr-ion with the stable 4f$^2$ state. Furthermore, $(c_{11}-c_{12})/2$ shows a crossover from the Curie-type softening proportional to the reciprocal temperature $1/T$ to a slight increase below 0.1 K. The latter increase in $(c_{11}-c_{12})/2$ below 0.1 K provides evidence of the quenching of the quadrupole $O_2^2$ of the non-Kramers $\Gamma_3$ doublet in PrInAg$_2$. From this result, one can deduce that quadrupolar Kondo scattering induces the quadrupolar Kondo state, where the octupole $T_{xyz}$ as well as the quadrupoles $O_2^0$ and $O_2^2$ are simultaneously quenched.

The low-temperature elastic constant $(c_{11}-c_{12})/2$ as a function of temperature in zero magnetic field is presented in Fig. 1(b). With decreasing temperature below 90 mK, $(c_{11}-c_{12})/2$ shows a slight increase proportional to $-\ln T$. This result is suggestive of the NFL behavior due to the overscreening effect in the quadrupolar Kondo state. However, the increase in the elastic constant $(c_{11}-c_{12})/2$ is different from the theoretical prediction,[22] which shows a logarithmic increase in quadrupolar susceptibility, namely, the softening of elastic constant. Our experimental result shows that quadrupolar susceptibility remains finite, suggesting that the quadrupolar moment of the non-Kramers $\Gamma_3$ doublet ground state is fully screened due to the quadrupolar Kondo effect below about 0.1 K. This discrepancy may arise from the coherent effect of the quadrupolar Kondo state in the periodic lattice of Pr ions.

To examine the magnetic field effect on the quadrupolar Kondo state, we measured the temperature dependence of $(c_{11}-c_{12})/2$ at various fields, and the results are shown in Fig. 2 (a). Figure 2 (b) shows the results of $(c_{11}-c_{12})/2$ calculated using Eqs. (1) and (2) in magnetic field. The experimental results in Fig. 2 (a) show that the magnetic field suppresses the softening of $(c_{11}-c_{12})/2$ at low temperatures. It is remarkable that at low magnetic fields below 5 T, the experimental results of $(c_{11}-c_{12})/2$ in Fig. 2 (a) deviate considerably from the calculation results in Fig. 2 (b), which is based on one-ion quadrupolar susceptibility for the stable $4f^2$ state with the non-Kramers $\Gamma_3$ doublet ground state. This means that the quadrupolar moment in the quadrupolar Kondo state is quenched at low magnetic fields below 5 T. At higher magnetic fields above 10 T, however, the experimental coincides with the calculation of one-ion quadrupolar susceptibility. This result suggests that high magnetic fields above 10 T collapse the quadrupolar Kondo state and consequently bring back the conventional paramagnetic state, which is well described in terms of one-ion quadrupolar susceptibility for the stable $4f^2$ state.

It is worthwhile to show the magnetic field dependence of $(c_{11}-c_{12})/2$ at a base temperature of 20 mK for the demonstration of the collapse of the quadrupolar Kondo state in magnetic fields. The field dependence of $(c_{11}-c_{12})/2$ at 20 mK is presented in Fig. 3 together with the field dependence of $(c_{11}-c_{12})/2$ calculated on the basis of one-ion quadrupolar susceptibility. The applied magnetic fields split the non-Kramers $\Gamma_3$ ground state doublet by hybridizing with the excited states. The quadrupolar Kondo state is collapsed when the Zeeman splitting energy $\Delta$ of the non-Kramers $\Gamma_3$ ground state exceeds the Kondo energy $k_B T_K$. Yatskar et al. estimated the Kondo temperature $T_K$ of PrInAg$_2$ to be ~0.86 K from the specific heat data.[14] Using the CEF parameters determined by the inelastic neutron scattering,[15] one can estimate the critical magnetic field $H_c$ = 8.7 T, where the Zeeman splitting energy becomes comparable to the Kondo energy $\Delta \sim k_B T_K$. Actually, the experimental results and calculations of $(c_{11}-c_{12})/2$ shown in Fig. 3 become very close to each other beyond the critical field $H_c$ = 8.7 T compatible for $\Delta \sim k_B T_K$, which is marked by a vertical dashed line. The considerable deviation between the experimental and calculation results of $(c_{11}-c_{12})/2$ at low magnetic fields below $H_c$ = 8.7 T down to a zero field is attributed to the quenching of the quadrupolar moment in the quadrupolar Kondo state.

Because the $\Gamma_3$ doublet of the Pr ion in the present PrInAg$_2$ arrays in a periodic lattice, the coherent effect of the quadrupolar Kondo state on the lattice might be possible in principle. In the present results, however, the behavior of $(c_{11}-c_{12})/2$ is mostly described within a framework of one-ion quadrupolar susceptibility and the impurity quadrupolar Kondo effect. No appreciable phenomenon indicating the coherent effect of the quadrupolar Kondo lattice is observed in the present experiments. Nevertheless, the heavy-Fermion quasi-particle resulting from the quadrupolar Kondo state for the doubly degenerate $\Gamma_3$ state is expected to possess an enhanced specific heat coefficient $\gamma = R\ln2/T_K = 6.7$ J/mol K$^2$ for $T_K \sim 0.86$ K. The experimental $\gamma = 7$ J/mol K$^2$ may promise the heavy-Fermion quasi-particle from the quadrupolar Kondo effect in PrInAg$_2$. The characteristic temperature $T^* = 1$ K $\sim 0.7$ K for Pr$_x$La$_{1-x}$InAg$_2$ reported by Kawae et al.[18] seems to be the Kondo temperature $T_K \sim 0.86$ K.

In conclusion, we have reported the low-temperature elastic properties of the crystalline PrInAg$_2$ possessing the non-Kramers $\Gamma_3$ doublet ground state. The elastic constant of $(c_{11}-c_{12})/2$ shows the softening proportional to the reciprocal temperature $1/T$ while lowering temperature down to 0.3 K, which is well reproduced by one-ion quadrupolar susceptibility indicating the stable $\Gamma_3$ ground state doublet in PrInAg$_2$. The temperature dependence of $(c_{11}-c_{12})/2$ above 10 T indicates again the stable character of the $\Gamma_3$ state at high magnetic fields. The suppression of the softening of $(c_{11}-c_{12})/2$ below 0.1 K at low magnetic fields below 5 T provides evidence of the quenching of the quadrupole $O_2^2$ in the quadrupolar Kondo state. The logarithmic behavior of $(c_{11}-c_{12})/2$ below 90 mK at a zero magnetic field is an important associated with the overscreening effect in the quadrupolar Kondo state in a periodic lattice. The quadrupolar Kondo effect of PrInAg$_2$ reported in the present letter will stimulate future research on strong correlation effect for the non-Kramers $\Gamma_3$ doublet interacting with conduction electrons.

We thank Prof. Y. Ōno of Niigata University. We thank staff members of the High Magnetic Field Laboratory of NIMS and the Center for Quantum Materials Science of Niigata University for their assistance. The present work was partially supported by a Grant-in-Aid for Scientific Research Priority Area (No. 15072202) of the Ministry of Education, Culture, Sports, Science, and Technology of Japan.

**Figure captions**

Fig.1

(a) Elastic constant $(c_{11}-c_{12})/2$ of PrInAg$_2$ as function of temperature at zero magnetic field. The solid line denotes the calculation results obtained using Eq. (1) based on quadrupolar susceptibility of Eq. (2). The quadrupole-strain interaction $|g_{\Gamma 3}| = 5.13$ K and intersite quadrupolar interaction g'$_{\Gamma 3}$ = - 0.004 K are used. The broken line denotes the background part of the elastic constant. The inset shows the low-temperature region of $(c_{11}-c_{12})/2$ on an expanded scale.

(b) Low-temperature region of elastic constant $(c_{11}-c_{12})/2$ as function of temperature at zero magnetic field. The solid line shows the $-\ln T$ dependence of $(c_{11}-c_{12})/2$ below 90 mK.

Fig. 2

Experimental (a) and calculation (b) results of elastic constant $(c_{11}-c_{12})/2$ as functions of temperature at various magnetic fields along [110] axis parallel to propagating direction of ultrasound wave. The coupling constant used in the calculation is the same one used in Fig. 1.

Fig.3

Elastic constant $(c_{11}-c_{12})/2$ as function of magnetic field at base temperature of 20 mK. The open circles and solid line in the inset denotes the experimental and calculation results, respectively. The broken line in the inset indicates the critical field $H_c$ = 8.7 T, at which the Zeeman splitting energy becomes equal to the Kondo temperature $T_K$ = 0.86 K after specific heat measurement[14].

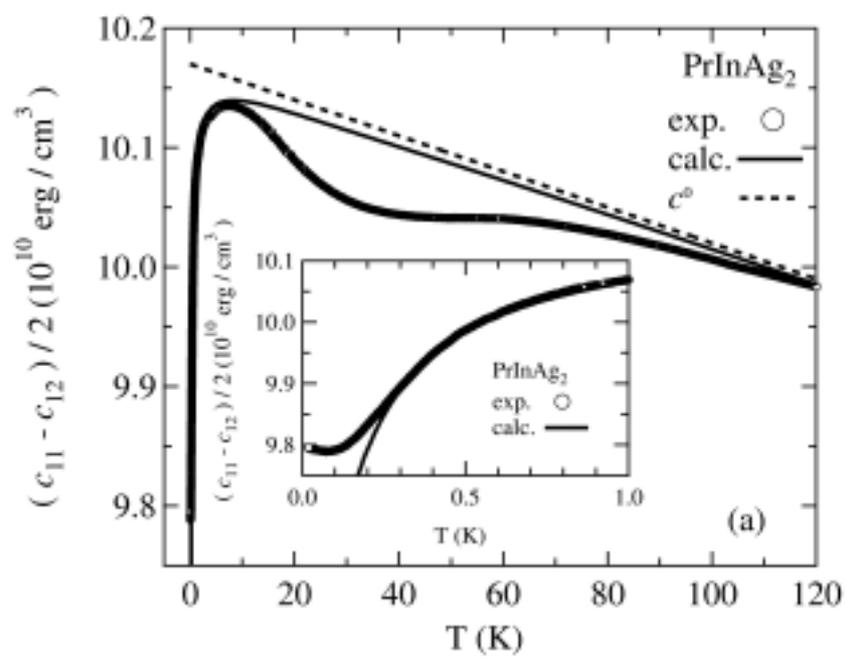

Fig. 1 (a).

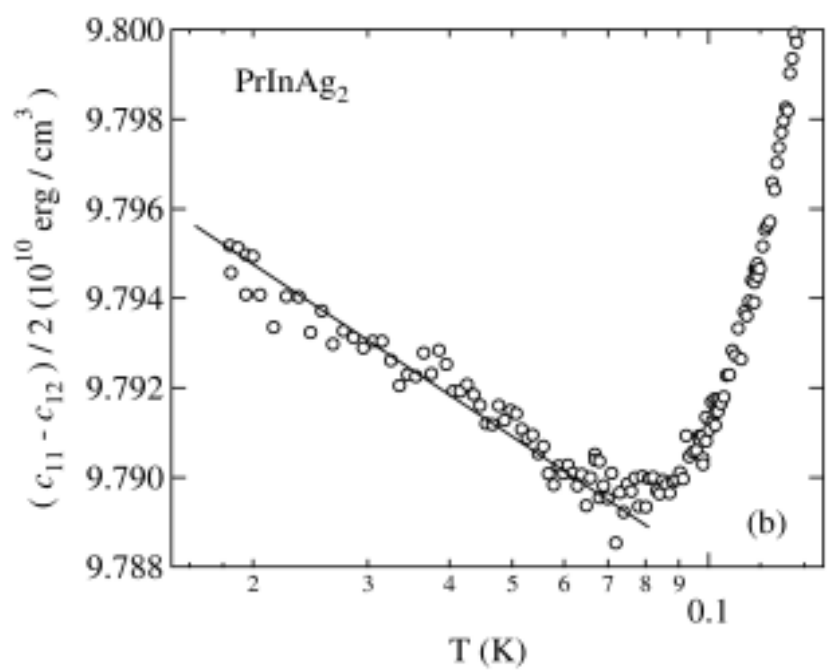

Fig. 1 (b).

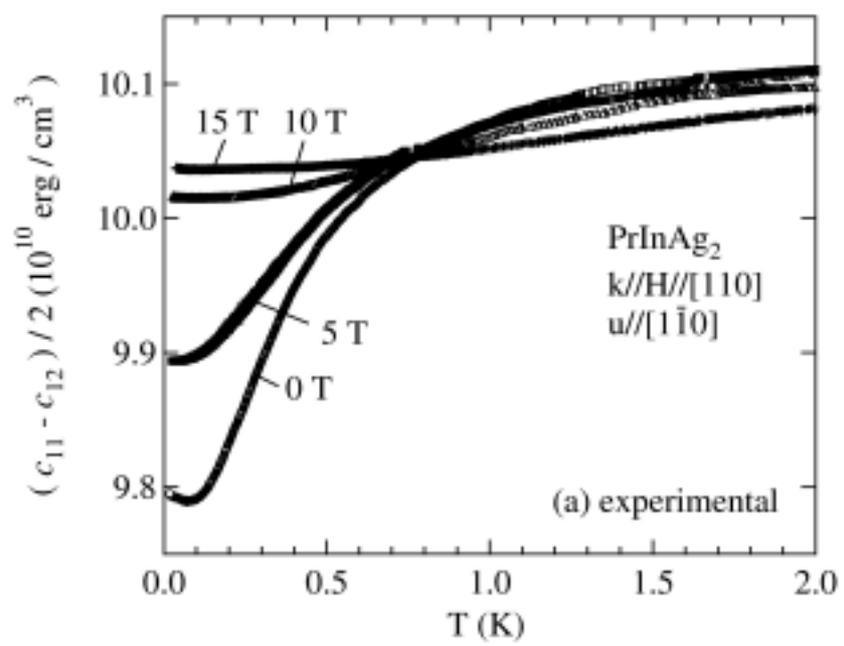

Fig.2 (a)

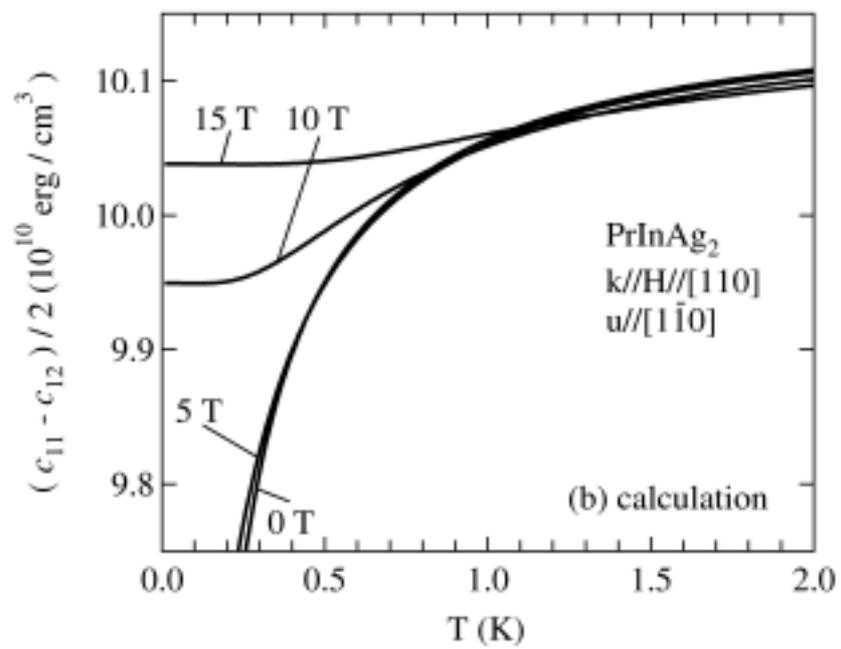

Fig.2 (b)

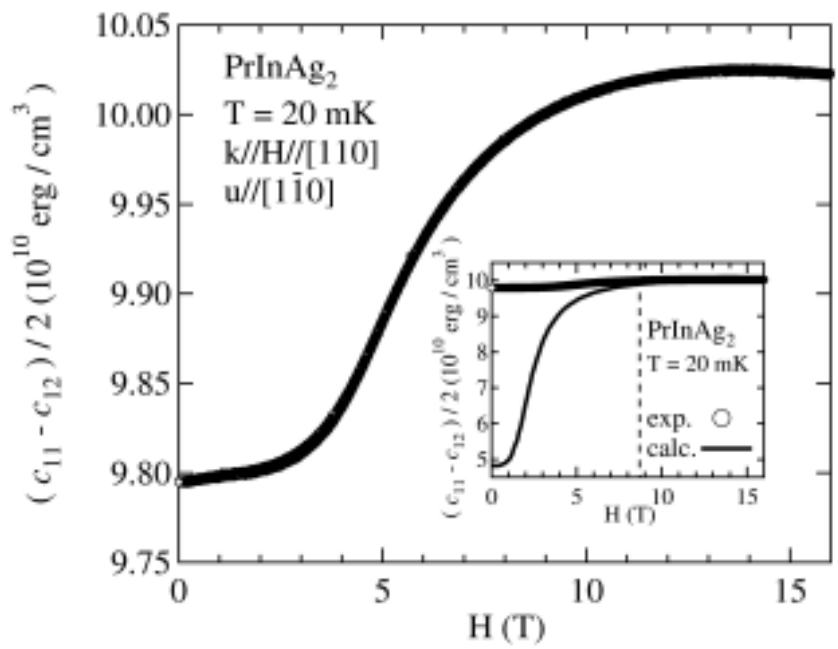

Fig. 3.